\documentclass[conference]{IEEEtran}
\usepackage{graphicx}
\usepackage{amsfonts}
\usepackage{amsmath,amsthm}
\usepackage{enumerate}
\usepackage{subfigure}

\ifCLASSINFOpdf
\else
\fi
\begin{document}
%
\title{Coalitional Game Theoretic Approach for Cooperative Transmission in Vehicular Networks}

\author{\IEEEauthorblockN{Tian Zhang\IEEEauthorrefmark{2}\IEEEauthorrefmark{1},
Wei Chen\IEEEauthorrefmark{2},
Zhu Han\IEEEauthorrefmark{3},
 and
Zhigang Cao\IEEEauthorrefmark{2}}
\IEEEauthorblockA{
\IEEEauthorblockA{\IEEEauthorrefmark{2}State Key Laboratory on Microwave
and Digital Communications,
\\
Tsinghua National Laboratory for Information Science and Technology
(TNList) \\
Department of Electronic Engineering, Tsinghua University, Beijing 100084, China
}
\IEEEauthorblockA{\IEEEauthorrefmark{1}School of Information Science and Engineering,
Shandong University, Jinan 250100, China }
\IEEEauthorrefmark{3}Department of Electrical and Computer Engineering,
University of Houston,
 Houston, TX 77004
\\ Email: tianzhang.ee@gmail.com, \{wchen, czgdee\}@tsinghua.edu.cn, zhan2@uh.edu}
}

\maketitle

\begin{abstract}
Cooperative transmission in vehicular networks is studied by using coalitional game and pricing in this paper. There are several vehicles and roadside units (RSUs) in the networks.
Each vehicle has a desire to transmit with a certain probability, which represents its data burtiness. The RSUs can enhance the vehicles' transmissions by cooperatively relaying the vehicles' data. We consider two kinds of cooperations: cooperation among the vehicles and cooperation between the vehicle and RSU. First, vehicles cooperate to avoid interfering transmissions by scheduling the transmissions of the vehicles in each coalition. Second, a RSU can join some coalition to cooperate the transmissions of the vehicles in that coalition. Moreover, due to the mobility of the vehicles, we introduce the notion of encounter between the vehicle and RSU to indicate the availability of the relay in space. To stimulate the RSU's cooperative relaying for the vehicles, the pricing mechanism is applied. A non-transferable utility (NTU) game is developed to analyze the behaviors of the vehicles and RSUs. The stability of the formulated game is studied. Finally, we present and discuss the numerical results for the 2-vehicle and 2-RSU scenario, and the numerical results verify the theoretical analysis.
\end{abstract}


%
\IEEEpeerreviewmaketitle
\section{Introduction}
Vehicular networks, from which drivers can obtain useful messages such as traffic conditions and real-time information on road to increase traffic safety and efficiency,
has gained much attention \cite{CST11: G. Karagiannis etc}. Meanwhile, vehicular networks can also provide entertainment content for passengers and collect data for road and traffic managers. In vehicular networks, vehicles and roadside units (RSUs) can communicate with each other through vehicle-to-vehicle (V2V), roadside-to-vehicle (R2V), vehicle-to-roadside (V2R) and roadside-to-roadside (R2R) communications.\par
As an important modus operandi of
substantially improving coverage and communication efficiency in wireless
networks, cooperative transmission has gained considerable attention recently.
In cooperative communications, some neighboring nodes can be used to relay the source signal to the destination, hence forming a virtual antenna array to obtain spatial diversity. Decode-and-forward (DF) is a commonly-used cooperative protocol \cite{IT04: J. N. Laneman D. N. C. Tse and G.W. Wornell}. In DF relaying, the relay node first decodes the received signal from the source, re-encodes it, and then forwards it to the destination. For the purpose of improving the system spectral efficiency,
the cooperative relaying with relay selection \cite{JSAC06:A. Bletsas A. Khisti D. P. Reed and A. Lippman,ICCC12: T. Zhang}
has been introduced on one hand. On the other hand, the non-orthogonal relaying protocols have been investigated \cite{JSAC04:R. U. Nabar H. Bolcskei and F.W. Kneubuhler}.
\par
Since the coalitional game theory provides analytical tools to model the behaviors of rational players when they cooperate, it is a powerful
tool for designing robust, practical, efficient, and fair cooperation strategies and has been extensively applied
in communication and wireless networks, which includes the vehicular networks \cite{SPM09:W. Saad Z. Han M. Debbah A. Hj ungnes and T. Basar}.
In \cite{JSAC08:S. Mathur L. Sankar and N. B. Mandayam}, the coalitional game theory was utilized to investigate the cooperation between rational wireless users, and the stability of the coalition was analyzed. Cooperative transmission between boundary nodes and backbone nodes was studied based on coalition games in \cite{TC09:Z. Han and H. Vincent Poor}.
 In \cite{WCNC10:D. Niyato P. Wang W. Saad and A. Hjphiungnes}, bandwidth sharing was studied by using coalition formation games in V2R communications.
In \cite{JSAC11:W. Saad Z. Han A. Hjphiungnes D. Niyato and E. Hossain}, the coalition formation games for distributed cooperation among RSUs in vehicular networks were studied. In \cite{TVT11:T. Chen F. Wu and S. Zhong}, the coalitional game theory was applied in studying how to stimulate message forwarding in vehicular networks.
The coalition formation problem for rational nodes in a cooperative DF
network was formulated in \cite{WCNC11:D. Niyato P. Wang W. Saad Z. Han and A. Hjphiungnes}.
The coalitional game theoretic approach for secondary spectrum access in cooperative cognitive radio networks was studied in \cite{TWC11:D. Li Y. Xu X. Wang and M. Guizani}. The stability of cooperation in multi-access systems was analyzed in \cite{ACCCC11:N. Karamchandani P. Minero M. Franceschetti}. In \cite{TMC12:K. Akkarajitsakul Ekram Hossain and Dusit Niyato}, the coalitional game was utilized to study the cooperative packet delivery in hybrid wireless networks.
In cooperative relay networks, there are costs (e.g., energy consumption, operational cost, bandwidth) at the relays for forwarding the other users' signals. Hence, a proper compensation mechanism is indispensable to provide the relays with incentives to forward the signals. Pricing mechanism was studied accordingly \cite{ICC06:N. Shastry and R. S. Adve}.
\par
In this paper, we investigate the cooperative transmission in vehicular networks under the framework of the coalitional game theory and pricing mechanism. On one hand, the vehicles can form coalitions to cooperatively schedule their transmissions. On the other hand, the RSUs can join the coalitions to cooperate the transmission of the vehicles. In the considered scenario, as the vehicles are dynamic with respect to the RSUs, the vehicles and RSUs may be very far away. In this case, the cooperative transmission may be unprofitable if not impossible.
Accordingly, we propose the notion of encounter between the vehicles and RSUs. Two conditions should be satisfied before a RSU cooperates a vehicle's transmission: 1) the RSU and vehicle are in the same coalition; 2) the RSU and vehicle encounter each other. When the two conditions hold, the RSU can use cooperative transmission to help the vehicle.
In return, the vehicle should pay for the RSU. We consider the problem as a non-transferable utility (NTU) game. The stability of the existing coalition is studied.
Numerical results demonstrate the efficiency of the proposed game.
\par
The reminder of the paper is structured as follows. Section II presents the system model. Next, the coalitional game approach for the considered problem is proposed in Section III. We first formulate a NTU game to model the cooperative transmission in the considered vehicular network, and then the analysis of the proposed coalitional game is carried out. In Section IV, the numerical results are discussed. Finally, we conclude the paper in Section V.


\section{System model}
\begin{figure}[]
\centering
\includegraphics[width=3.0in]{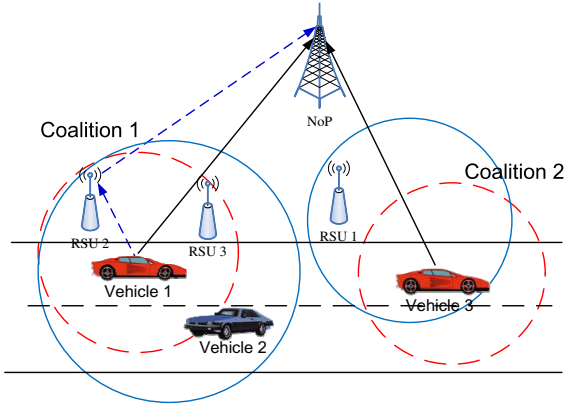}
\caption{Cooperative transmission with coalitions in vehicular networks}
\label{fig_system model}
\end{figure}

Consider a wireless network in Fig. \ref{fig_system model}, which consists of a network operator (NoP), $K$ vehicles, and $M$ RSUs.\footnote{The concept of relay node has been introduced in IEEE 802.16j for WiMAX networks.} The vehicles and RSUs can form coalitions and the RSUs can cooperate the transmissions of the vehicles when they are in the same coalition. Let $0$ denote the NoP, and $\mathbb{V}=\{1,2,\cdots,K\}$ and $\mathbb{R}=\{K+1,2,\cdots,K+M\}$ represent the set of the vehicles and RSUs, respectively.
\par
We consider the uplink communications from the vehicle to the NoP. Each $i \in  \mathbb{V}$ has its transmission range $d_i$.\footnote{It is assumed that within the transmission range, the correct decoding at the receiver can be guaranteed.}
 When RSU $i \in \mathbb{R}$ is in the transmission range of vehicles $j\in \mathbb{V}$, we call $i$ encounters $j$.
Vehicle $i \in \mathbb{V}$ is active
in each time-slot with probability $p_i$, independently of other vehicles.
When two or more vehicles transmit simultaneously, it is called a collision, and we suppose that the transmissions will fail when a collision occurs.
The cooperative protocol utilized in this paper is the non-orthogonal decode-and-forward (NDF) \cite{JSAC04:R. U. Nabar H. Bolcskei and F.W. Kneubuhler}. In the first half of the time-slot, the source transmit to the relay and destination; In the second half, the relay decodes and forwards the source message. Meanwhile, the source transmits a new message to the destination.

We assume that vehicles in the same coalition can cooperate to schedule the transmissions, and only one vehicle transmits at a time to avoid interference. In each coalition, a scheduler determines the active user that can transmit while other active users remain silent. When vehicle $j$ is scheduled to transmit at a time-slot,
it selects one RSU from the feasible RSU set (i.e., the set of RSUs that are in the same coalition with vehicle $j$ and encounter vehicle $j$) as relay to assist its transmission.\footnote{We assume that the selection is performed according to a uniform probability distribution for simplicity.} Then one vehicle only employs at most one RSU as relay for each transmission. Furthermore, as no more than one vehicle is scheduled to transmit at a given time-slot, one RSU assists the most one vehicles at a given time-slot.
When the vehicle $j$ utilizes RSU $i$ as relay for its transmission, RSU $i$ charges vehicle $j$ with price $\xi_{ij}$ per transmission.
\par
When some vehicles and RSUs form a coalition, the vehicles would share the channel with each other in TDMA mode, and they need to pay for the RSU's relaying. However, the vehicles can avoid collision and gain diversity as well as rate increase. On the other hand, although there are costs in receiving and forwarding the vehicles' signals \cite{TWC09:W. Saad Z. Han M. Debbah and A. Hjphiungnes}, the RSUs could achieve revenues by charging the vehicles. In a word, both vehicles and RSUs have incentives to form coalitions.
\section{Coalitional game approach}
In this section, we first formulate the coalitional game in Section \ref{Formulation}, and then we analyze the formulated game in Section \ref{Analysis}.
\theoremstyle{definition} \newtheorem{defination}{Defination}
\subsection{Coalitional game formulation}\label{Formulation}
A coalitional game $\mathcal{G}$ is
uniquely defined by the pair $(\mathcal{N},v)$, where $\mathcal{N}$ is the set of players, any non-empty subset $S \subseteq \mathcal{N}$ is called a coalition, and $v$ is the coalition value,
it quantifies the worth of a coalition in a game.
\par
In our paper, the players are the vehicles and RSUs, i.e., $\mathcal{N}=\mathbb{V}\cup \mathbb{R}$. $S \subseteq \mathcal{N}$ is a coalition. Define $S\cap\mathbb{V}:=S_u$ and $S\cap\mathbb{R}:=S_r$.
Consider a time-slot, let $i\leftrightarrow j$ denote $i$ encounters $j$ during the whole time-slot and $\mathcal{P}_{ij}=\Pr\left\{i \leftrightarrow j\right\}$.
The data rate increase of vehicle $i$ with the cooperation from the RSU $j$ is denoted as $\Delta_{ij}$.
Formally, the scheduler in $S$ is a map $f_S: 2^{S} \to S$ such that $f_S(\Psi) \in \Psi$ for all $\Psi \subseteq S$ and $f_S(\Psi)=\emptyset$ iff $\Psi=\emptyset$.\footnote{Similar definitions can be found in \cite{ACCCC11:N. Karamchandani P. Minero M. Franceschetti}.}

The average effective throughput for vehicle $i$ can be expressed as $$T_i(S)=\mathbb{E}_{\Psi}\left\{\textbf{1}_{f_S(\Psi)=\{i\}}\right\}\left(1+\zeta_i(S)\right)\prod\limits_{j\in \mathbb{V}\backslash S_u}\left(1-p_j\right),$$
where
$\mathbb{E}_{\Psi}\left\{\textbf{1}_{f_S(\Psi)=\{i\}}\right\}$ denotes the ratio of time-slots that vehicle $i$ is chosen to transmit. For example, we can assume $f_S(\Psi)$ chooses the minimal element from the set of active vehicles $\Psi$.
Let $S_u=\left\{s_1,\cdots,s_{|S_u|}\right\}$ with $s_1>\cdots >s_{|S_u|}$, then
\begin{eqnarray}\label{ratio for the minimal elemetn in Su}
\mathbb{E}_{\Psi}\left\{\textbf{1}_{f_S(\Psi)=\{s_{|S_u|}\}}\right\}=p_{s_{|S_u|}}
\end{eqnarray}
 and
 \begin{eqnarray}\label{ratio for k-th lagrest the elemetn in Su}
 \lefteqn{
 \mathbb{E}_{\Psi}\left\{\textbf{1}_{f_S(\Psi)
=\{s_{k}\}}\right\}
 }
 \nonumber \\
&=&p_{s_{k}}\prod\limits_{i=k+1}^{|S_u|}(1-p_{s_i}), k=1,\cdots,|S_u|-1.
  \end{eqnarray}
$\zeta_i(S)$ is the average increase of data rate for vehicle $i$ and it is given by
\begin{eqnarray}
\lefteqn{
\zeta_i(S)=\sum\limits_{j\in S_r}\Delta_{ij}\mathcal{P}_{ji}\prod\limits_{k \ne j\in S_r}(1-\mathcal{P}_{ki})
}
\nonumber \\
&+&\sum\limits_{j<k  \in S_r}\frac{\Delta_{ij}+\Delta_{ik}}{2}\mathcal{P}_{ji}\mathcal{P}_{ki}\prod\limits_{l \ne j,l \ne k \in S_r}(1-\mathcal{P}_{li})\nonumber \\
&+&\sum\limits_{j<k <l\in S_r}\frac{\Delta_{ij}+\Delta_{ik}+\Delta_{il}}{3}\mathcal{P}_{ji}\mathcal{P}_{ki}\mathcal{P}_{li}\nonumber \\
&\times&\prod\limits_{r \ne j,r \ne k,r\ne l \in S_r}(1-\mathcal{P}_{ri})
+
\cdots
+\frac{\sum\limits_{j\in S_r}\Delta_{ij}}{n}\prod\limits_{j\in S_r}\mathcal{P}_{ji}, \nonumber
\end{eqnarray}
where $n=|S_r|$. If $S_r=\emptyset$, i.e., $n=0$, $\zeta_i(S)=0$.
Suppose $\Delta_{ij}=\Delta_{i}$, i.e., the data increase for vehicle $i$ is irrelevant to the selection of RSUs, then the average throughput for vehicle $i$ can be simplified as
$
T_i(S)=\mathbb{E}_{\Psi}\left\{\textbf{1}_{f_S(\Psi)=\{i\}}\right\}\left(1+\mathcal{P}_{i}\Delta_{i}\right)\prod\limits_{j\in \mathbb{V}\backslash S_u}\left(1-p_j\right),
$
where
$\mathcal{P}_{i}=1-\prod\limits_{j\in S_r}\left(1-\mathcal{P}_{ji}\right)$ is the probability that at least one RSU in the coalition encounters vehicle $i$.

\emph{Remark: If there are active vehicles outside $S$ at a given time-slot, at least one vehicle outside $S$ transmits simultaneously with the scheduled vehicles in $S$ no matter how the vehicles outside $S$ form coalitions. Thus, there is no collision if and only if (iff) all vehicles outside $S$ is inactive. }

For vehicle $i$, the average payment made to the RSUs can be given by
$
P_i(S)=\mathbb{E}_{\Psi}\left\{\textbf{1}_{f_S(\Psi)=\{i\}}\right\}\chi_i(S)
$
with
\begin{eqnarray}
\lefteqn{
\chi_i(S)=
\sum\limits_{j\in S_r}\xi_{ji}\mathcal{P}_{ji}\prod\limits_{k \ne j\in S_r}(1-\mathcal{P}_{ki})
}
\nonumber \\
&+&\sum\limits_{j<k  \in S_r}\frac{\xi_{ji}+\xi_{ki}}{2}\mathcal{P}_{ji}\mathcal{P}_{ki}\prod\limits_{l \ne j,l \ne k \in S_r}(1-\mathcal{P}_{li})\nonumber \\
&+&\sum\limits_{j<k <l\in S_r}\frac{\xi_{ji}+\xi_{ki}+\xi_{li}}{3}\mathcal{P}_{ji}\mathcal{P}_{ki}\mathcal{P}_{li}\nonumber \\
&\times&\prod\limits_{r \ne j,r \ne k,r\ne l \in S_r}(1-\mathcal{P}_{ri})
+
\cdots
+\frac{\sum\limits_{j\in S_r}\xi_{ji}}{n}\prod\limits_{j\in S_r}\mathcal{P}_{ji}.\nonumber
\end{eqnarray}
If $n=0$, $\chi_i(S)=0$.
Assume $\xi_{ij}=\xi_j$, i.e., all the RSUs set the same price for vehicle $j$, then
the average payment can be simplified as
$
P_i(S)=\mathbb{E}_{\Psi}\left\{\textbf{1}_{f_S(\Psi)=\{i\}}\right\}\mathcal{P}_{i}\xi_i.
$
\par
\emph{Remark: The RSU charges the vehicles once the vehicles employs the RSU as the relay for a transmission, it does not take the collisions into account.
That is to say, the actions in other coalitions do not affect the charging.}
\par
The payoff of vehicle $i$ is determined by
$
u_i(S)=\alpha_i T_i(S)-\beta_iP_i(S).
$
\par
For RSU $j$ in $S$, the revenue charged from the vehicles can be given by
\begin{eqnarray}\label{R}
R_j(S)=\sum\limits_{i\in S_u}\mathbb{E}_{\Psi}\left\{\textbf{1}_{f_S(\Psi)=\{i\}}\right\}\eta_{ij}(S)\xi_{ji},
\end{eqnarray}
where $\eta_{ij}(S)$ is the probability that vehicle $i$ employs RSU $j$ as its relay for transmission, and it is given by
\begin{eqnarray}
\eta_{ij}(S)&=&\mathcal{P}_{ji}
\bigg[\prod\limits_{k \ne j\in S_r}(1-\mathcal{P}_{ki})\nonumber \\
&+&\frac{1}{2}\sum\limits_{k \ne j\in S_r}\mathcal{P}_{ki}\prod\limits_{l \ne j,l \ne k \in S_r}(1-\mathcal{P}_{li})\nonumber \\
&+&\frac{1}{3}\sum\limits_{k <l, k\ne j, l\ne j\in S_r}\mathcal{P}_{ki}\mathcal{P}_{li}\prod\limits_{r \ne j,r \ne k,r\ne l \in S_r}(1-\mathcal{P}_{ri})\nonumber \\
&+&
\cdots
+\frac{1}{n}\prod\limits_{k \ne j\in S_r}\mathcal{P}_{ki}\bigg].
\end{eqnarray}
\par
Assume that RSU $j$ receives the signal of vehicle $i$ at cost $c_{ji}^{r}$ and the cost of forwarding the signal to NoP as $c_{ji}^{f}$.
The average cost of RSU $j$ can be expressed as
\begin{eqnarray}\label{C}
C_j(S)=\sum\limits_{i\in S_u}\mathbb{E}_{\Psi}\left\{\textbf{1}_{f_S(\Psi)=\{i\}}\right\}\Big[c_{ji}^{f}\eta_{ij}(S)+\mathcal{P}_{ji}c_{ji}^{r}\Big].
\end{eqnarray}
\par
\emph{Remark: RSU $j$ receives the message of vehicle $i$ once it encounters vehicle $i$ (with probability $\mathcal{P}_{ji}$), and it forwards the message only when it is selected as the relay by vehicle $i$ (with probability $\eta_{ij}(S)$).}\par
The payoff of RSU $j$ is determined by
$\tilde{u}_j(S)=\gamma_j R_j(S)-\mu_jC_j(S).$
Define $v(S) \subseteq \mathfrak{R}^{|S_u|+|S_r|}$ be the set of feasible payoff vectors for $S$,
we formulate the considered cooperation problem as a coalitional NTU-game $\mathcal{G}:(\mathbb{V}\cup \mathbb{R},v)$.
\subsection{Analysis of the formulated coalitional game}\label{Analysis}
In this section, we first present two observations. Next, we analyze the stability of the game and propose a sufficient condition for the existence of the core.
\theoremstyle{definition} \newtheorem{observation}{Observation}
\par
In the beginning, we have the following observation.
\begin{observation}\label{Cancellation}
Let
$
f(S)=\sum\limits_{i \in S_u} u_i(S)+\sum\limits_{j \in S_r} \tilde{u}_j(S)
$
denote the sum payoff of $S$.
When $\gamma_j=1$ and $\beta_i=1$, we have $f(S)=\sum\limits_{i \in S_u} \alpha_iT_i(S)-\sum\limits_{j \in S_r} \mu_jC_j(S)$. That is to say, the pricing has no effect on the sum payoff in this case.
\end{observation}
\begin{IEEEproof}
First we can prove that
$
\chi_i=\sum\limits_{j \in S_r}\eta_{ij}(S)\xi_{ji}.
$
Then
\begin{eqnarray}\label{1}
\lefteqn{
\mathbb{E}_{\Psi}\left\{\textbf{1}_{f_S(\Psi)=\{i\}}\right\}\chi_i
}
\nonumber \\
&=&
\mathbb{E}_{\Psi}\left\{\textbf{1}_{f_S(\Psi)=\{i\}}\right\}\sum\limits_{j \in S_r}\eta_{ij}(S)\xi_{ji}
\nonumber \\
&\stackrel{(a)}{=}& \sum\limits_{j \in S_r}\mathbb{E}_{\Psi}\left\{\textbf{1}_{f_S(\Psi)=\{i\}}\right\}\eta_{ij}(S)\xi_{ji}.
\end{eqnarray}
($a$) holds since $\mathbb{E}_{\Psi}\left\{\textbf{1}_{f_S(\Psi)=\{i\}}\right\}$ is irrelevant to $j \in S_r$.
Next, based on (\ref{1}), we can derive
\begin{eqnarray}
\lefteqn{
\sum\limits_{i\in S_u}\mathbb{E}_{\Psi}\left\{\textbf{1}_{f_S(\Psi)=\{i\}}\right\}\chi_i
}
\nonumber \\
&= & \sum\limits_{i\in S_u}\sum\limits_{j \in S_r}\mathbb{E}_{\Psi}\left\{\textbf{1}_{f_S(\Psi)=\{i\}}\right\}\eta_{ij}(S)\xi_{ji}.
\end{eqnarray}
Exchanging the summation order on the right side, we get
\begin{eqnarray}\label{preProof}
\sum\limits_{i \in S_u} P_i(S)=\sum\limits_{j \in S_r} R_j(S).
\end{eqnarray}
When $\gamma_j=1$ and $\beta_i=1$,
$
f(S)=\sum\limits_{i \in S_u} \alpha_iT_i(S)-\sum\limits_{j \in S_r} \mu_jC_j(S)
+\sum\limits_{i \in S_u} P_i(S)-\sum\limits_{j \in S_r} R_j(S).
$
Using (\ref{preProof}), we derive $f(S)=\sum\limits_{i \in S_u} \alpha_iT_i(S)-\sum\limits_{j \in S_r} \mu_jC_j(S)$.
\end{IEEEproof}
\emph{Remark: Observation \ref{Cancellation} reveals the fact that the total revenues obtained by the RSUs equal to the payments of all the vehicles. }
\par
In addition, we obtain the second observation.
\begin{observation}\label{RNnocoalition}
A coalition $S$ should have at least one vehicle. Otherwise, $\tilde{u}_i(S)=0=\tilde{u}_i(\{i\})$ and $v(S)=\sum\limits_{i \in S}\tilde{u}_i(S)=0$. That is to say, when there are only the RSUs, the RSUs have no stimuli to form coalitions and each RSU will act alone.
\end{observation}
\begin{IEEEproof}
When $|S_u|=0$, we get $R_j(S)=0$ and $C_j(S)=0$ according to (\ref{R}) and (\ref{C}), respectively. Thus, $\tilde{u}_i(S)=0$. Specifically, $\tilde{u}_i(\{i\})=0$. As $\tilde{u}_i(S)=\tilde{u}_i(\{i\})$ in this case, each RSU will act alone.
\end{IEEEproof}
\emph{Remark: The function of the RSU is relaying the vehicle's signal. So when there is no vehicle, it is meaningless to group only the RSUs together.}
\par
On the other hand,
when there is no RSU in a coalition $S$, i.e., $S \subseteq \mathbb{V}$, $u_i(S)=\mathbb{E}_{\Psi}\left\{\textbf{1}_{f_S(\Psi)=\{i\}}\right\}\prod\limits_{j\in \mathbb{V}\backslash S}\left(1-p_j\right)$ for $i \in S$ and
$v(S)=\sum\limits_{i \in S}u_i(S)>0$. Specially when $S=\{i\}$, we derive $u_i(\{i\})=p_i\prod\limits_{j\in \mathbb{V} \backslash \{i\}}\left(1-p_j\right)$. Hence, when $\exists S \subseteq \mathbb{V}~\&~ S \owns i$ satisfying $u_i(S)>u_i(\{i\})$, the vehicles will form coalitions to improve the utility. Specifically, let $S=\{s_1,\cdots,s_{|S|}\}$ with $s_1>\cdots>s_{|S|}$, based on (\ref{ratio for the minimal elemetn in Su}) and (\ref{ratio for k-th lagrest the elemetn in Su}), we can derive that if
\begin{eqnarray}\label{condition for forming coalition for pure vehicle}
1 \ge \left\{
\begin{array}{ll}
\prod\limits_{j \in S\backslash \{s_i\}}\left(1-p_j\right), &  i=|S|;\\
\frac{\prod\limits_{j \in S\backslash \{s_i\}}\left(1-p_j\right)}
{\prod\limits_{k =i+1}^{|S|}\left(1-p_{s_k}\right)} , &  \mathrm{otherwise},
\end{array} \right.
\end{eqnarray}
forming coalition $S$ is profitable.\footnote{Although forming $S$ may be not optimal, it is at least better than acting alone.} Specially, when $p_i=p$, i.e., all vehicles have the same active probability, we can derive that (\ref{condition for forming coalition for pure vehicle}) holds, then forming coalitions is always profitable in the case.
\par
Next, as the core is one of the most important stability
concepts defined for coalitional games, we investigate the core of our proposed coalitional game in the following.
\par
The definition for the core of our coalitional game is given as follows.
\begin{defination}
The core of $(\mathbb{V}\cup \mathbb{R},v)$ is defined as $C=\big\{x\in  v\left(\mathbb{V}\cup \mathbb{R}\right): \forall S, \not\exists y \in v(S), s.t. ~ y_i>x_i, \forall i \in S\big\}$.
\end{defination}
\par
The following observation gives a sufficient condition for the existence of the core.
\begin{observation}\label{sufficient condition for the existence of the core}
The core of $(\mathbb{V}\cup \mathbb{R},v)$ is nonempty once the following conditions hold ($S \subset \mathbb{V}\cup \mathbb{R}$):
\begin{enumerate}[1)]
\item $\alpha_i>0$, $\beta_i>0$, $\gamma_j>0$, and $\mu_j>0$.
\item  $\alpha_i T_i(S)>\beta_iP_i(S)$ or $ \gamma_j R_j(S)>\mu_jC_j(S)$.
\item $ \alpha_i T_i(\mathbb{V}\cup \mathbb{R})-\beta_iP_i(\mathbb{V}\cup \mathbb{R})>\alpha_i T_i(S)-\beta_iP_i(S)$, and $\gamma_j R_j(\mathbb{V}\cup \mathbb{R})-\mu_jC_j(\mathbb{V}\cup \mathbb{R})>\gamma_j R_j(S)-\mu_jC_j(S)$.
\end{enumerate}
\end{observation}
\begin{IEEEproof}
When 1) holds, we can find $\alpha_i$, $\beta_i$, $\gamma_i$, and $\mu_j$ to satisfy 2).  If 2) does not holds,
we have
$u_i(\{i\})=\prod\limits_{j\in \mathbb{V}/\{i\}}\left(1-p_j\right)\ge 0\ge u_i(S)$ for $i \in S_u$
and $\tilde{u}_j(S)\le 0=\tilde{u}_j(\{j\})$ for $j \in S_r$. Then, each vehicle and RSU will act alone. In this case, the core is empty. When 3) holds, we can prove that $(\mathbb{V}\cup \mathbb{R},v)$ is balanced \cite{Book91:R. B. Myerson}. Thus, the core is nonempty according to the Bondareva-Shapley theorem \cite{Book01:Zhu Han Dusit Niyato Walid Saad Tamer Basar and Are Hjorungnes}.
\end{IEEEproof}
\emph{Remark: The core is possibly non-empty in practice. For example, when the considered vehicles wait for the traffic light,
the vehicles as well as the nearby RSUs are probable to form the coalition together.}
\section{Numerical results}
In this section, we demonstrate the numerical evaluations for the performance of the cooperative transmission scheme with coalitions.
In the simulations, we assume that the nodes are uniformly located in a square area of $1km \times 1km$.\footnote{The area has been divided to $10 \times 10$.} The network topology changes at the beginning of each time-slot and remains static during the whole time-slot. That is to say, the locations of the nodes are generated according to the uniform distribution at the beginning of a time-slot, the locations do not change during the time-slot, and we re-generate the locations at the beginning of the next time-slot.
We consider the scenario that there are 2 vehicles (vehicle 1 and vehicle 2) and 2 RSUs (RSU 3 and RSU 4) in the area.
\begin{figure}[h]
\begin{center}
\includegraphics[width=2.6in]{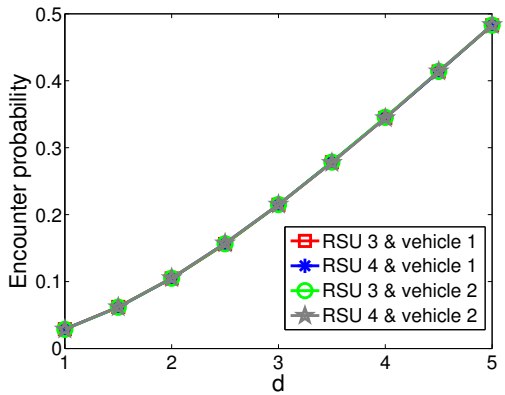}
\caption{Encounter probability with different transmission ranges}
\label{encounterprob}
\end{center}
\end{figure}
\par
Fig. \ref{encounterprob} shows the encounter probability with different transmission ranges for vehicle 1 and vehicle 2. In the simulations, we set $d_1=d_2=d$ and the probability is obtained from $10^6$ time-slots. We can observe that RSU 3 \& vehicle 1, RSU 4 \& vehicle 1, RSU 3 \& vehicle 2, and RSU 4 \& vehicle 2 have similar encounter probabilities.
It is because that since all nodes are uniformly distributed, vehicle 1 and vehicle 2 as well as RSU 3 and RSU 4 are exchangeable in location. In addition, we can see that the encounter probability increases with the increase of the transmission range.
\begin{table}[]
\caption{Possible coalitional structure}
\label{tab:1}
\centering
\begin{tabular}{|l|l|l|l|}
  \hline
   $\mathcal{C}_1$: \{1,2,3,4\}& $\mathcal{C}_6$: \{1,3\},\{2,4\} &   $\mathcal{C}_{11}$: \{1,2\},\{3,4\}\\
   $\mathcal{C}_2$: \{1,3,4\},\{2\}& $\mathcal{C}_7$: \{1,2,3\},\{4\} &  $\mathcal{C}_{12}$: \{1\},\{2\},\{3,4\}  \\
   $\mathcal{C}_3$: \{1,2\},\{3\},\{4\}& $\mathcal{C}_8$: \{1\},\{2,3,4\} &  $\mathcal{C}_{13}$: \{1,2,4\},\{3\}  \\
  $\mathcal{C}_4$: \{1\},\{2\},\{3\},\{4\} & $\mathcal{C}_9$: \{1,4\},\{2,3\} &    $\mathcal{C}_{14}$: \{1,4\},\{2\},\{3\} \\
  $\mathcal{C}_5$: \{1\},\{3\},\{2,4\} & $\mathcal{C}_{10}$: \{1\},\{4\},\{2,3\}  &   $\mathcal{C}_{15}$: \{2\},\{4\},\{1,3\}\\
  \hline
\end{tabular}
\end{table}
In the evaluations of utility performance, we set $\Delta_{ij}=0.5$, $p_i=0.6$, $\xi_{ij}=1.5$, $c_{ji}^{f}=0.5$, $c_{ji}^{r}=0.2$, $\alpha_i=10$, $\beta_i=1$, and $\gamma_j=\mu_j=1$. There are totally 15 coalitional structures for 2 vehicles and 2 RSUs as illustrated in Table \ref{tab:1}. Using Observation \ref{RNnocoalition}, we need not considering $\mathcal{C}_{11}$ and $\mathcal{C}_{12}$. Meanwhile,
as vehicle 1 and vehicle 2 as well as RSU 3 and RSU 4 are exchangeable,\footnote{vehicle 1 and vehicle 2 are not exchangeable because of the scheduling when they are in the same coalition. However, as exchanging elements in the same coalition is meaningless, it does not affect the analysis here.}
we only need to consider $\mathcal{C}_1$ - $\mathcal{C}_7$.\footnote{$\mathcal{C}_8$ is similar as $\mathcal{C}_2$; $\mathcal{C}_9$ is similar as
$\mathcal{C}_6$; $\mathcal{C}_{10}$, $\mathcal{C}_{14}$ and $\mathcal{C}_{15}$ are similar as $\mathcal{C}_5$; and $\mathcal{C}_{13}$ is similar as $\mathcal{C}_7$.}
\begin{figure}[!t]
\begin{center}
\includegraphics[width=2.6in]{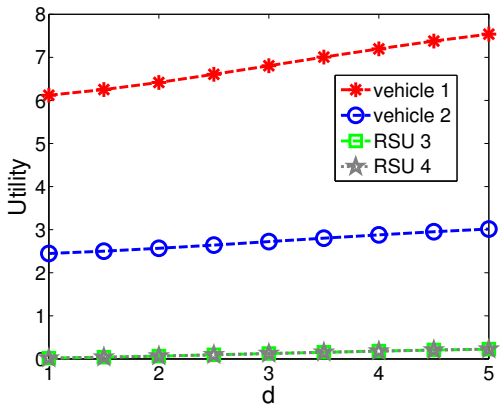}
\caption{Utility performance in $\mathcal{C}_1$}
\label{C1}
\end{center}
\end{figure}
\par
Fig. \ref{C1} plots the utility performance for 4 nodes when they form the coalition together($\mathcal{C}_1$). The utility performance increases when we increase the transmission range. The reason is that the encounter probability will increase when the transmission range increases (see Fig. \ref{encounterprob}). Consequently, the probability of cooperative transmission will increase. As cooperative transmission could benefit both the vehicle and RSU,\footnote{Under the simulation settings, cooperative transmission is preferable for both vehicle and RSU. } the utility performance increases. Another observation is that the utility performance for vehicle 1 is better than that of vehicle 2, and the utility performance for RSU 3 and RSU 4 is similar. This can be explained as follows: when vehicle 1 and vehicle 2 are in the same coalition and both of them are active, the scheduler selects vehicle 1 to transmit,
i.e., the vehicle 1 has higher transmission priory than vehicle 2. Consequently, the utility performance for vehicle 1 is better. In contrast, RSU 3 and RSU 4 have same priority in the relay selection of vehicle and they have same encounter probability with vehicle 1 (or vehicle 2), the same relaying price, and the same cost for receiving and forwarding,
so they have similar utility performance.
\begin{figure}[!t]
\begin{center}
\subfigure[$\mathcal{C}_2$]{\label{C2}\includegraphics[width=1.7in]{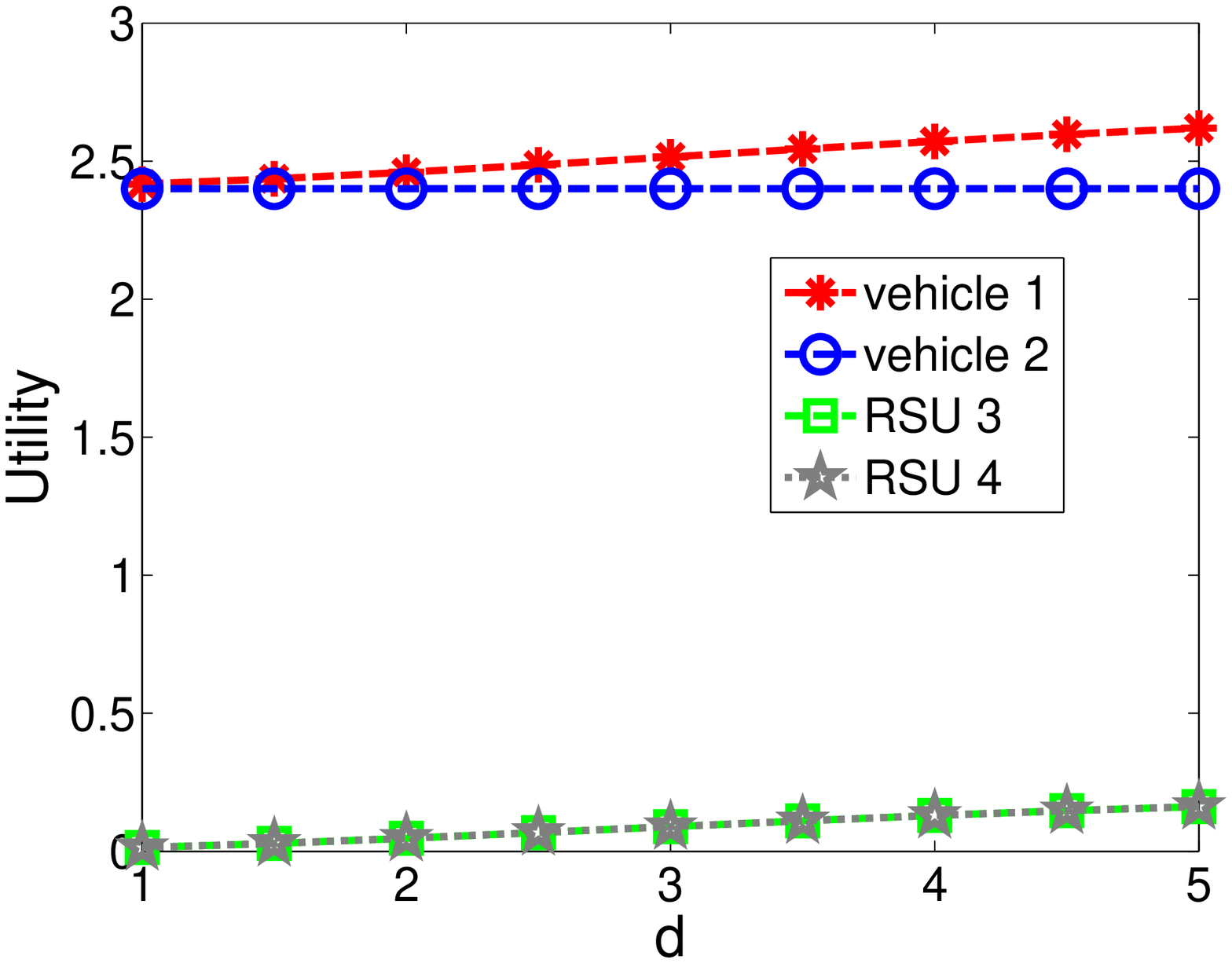}}
\subfigure[$\mathcal{C}_3$]{\label{C3}\includegraphics[width=1.7in]{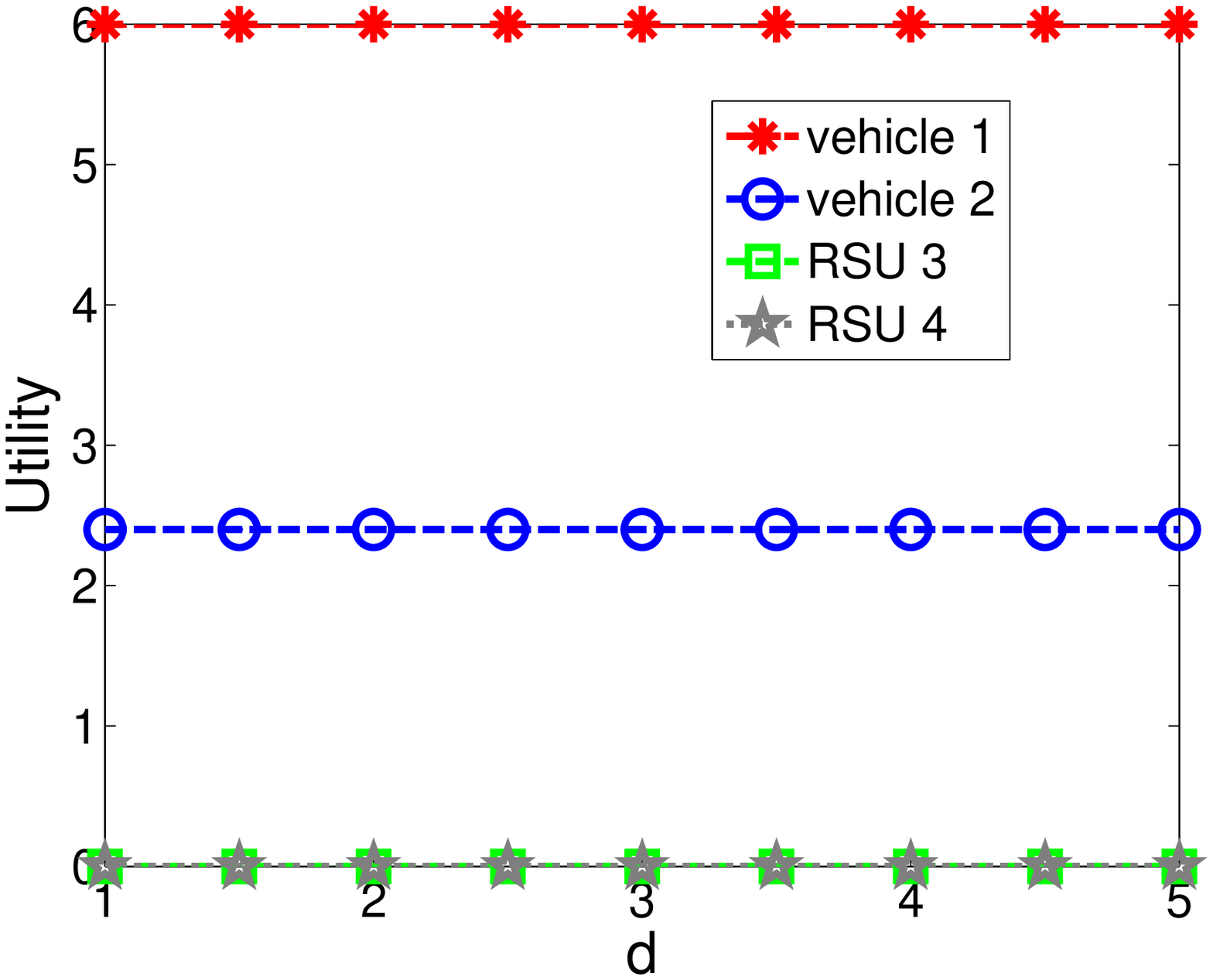}}
\caption{Utility performance in $\mathcal{C}_2$ and $\mathcal{C}_3$}
\end{center}
\end{figure}
\begin{figure}[!t]
\begin{center}
\subfigure[$\mathcal{C}_4$]{\label{C4}\includegraphics[width=1.7in]{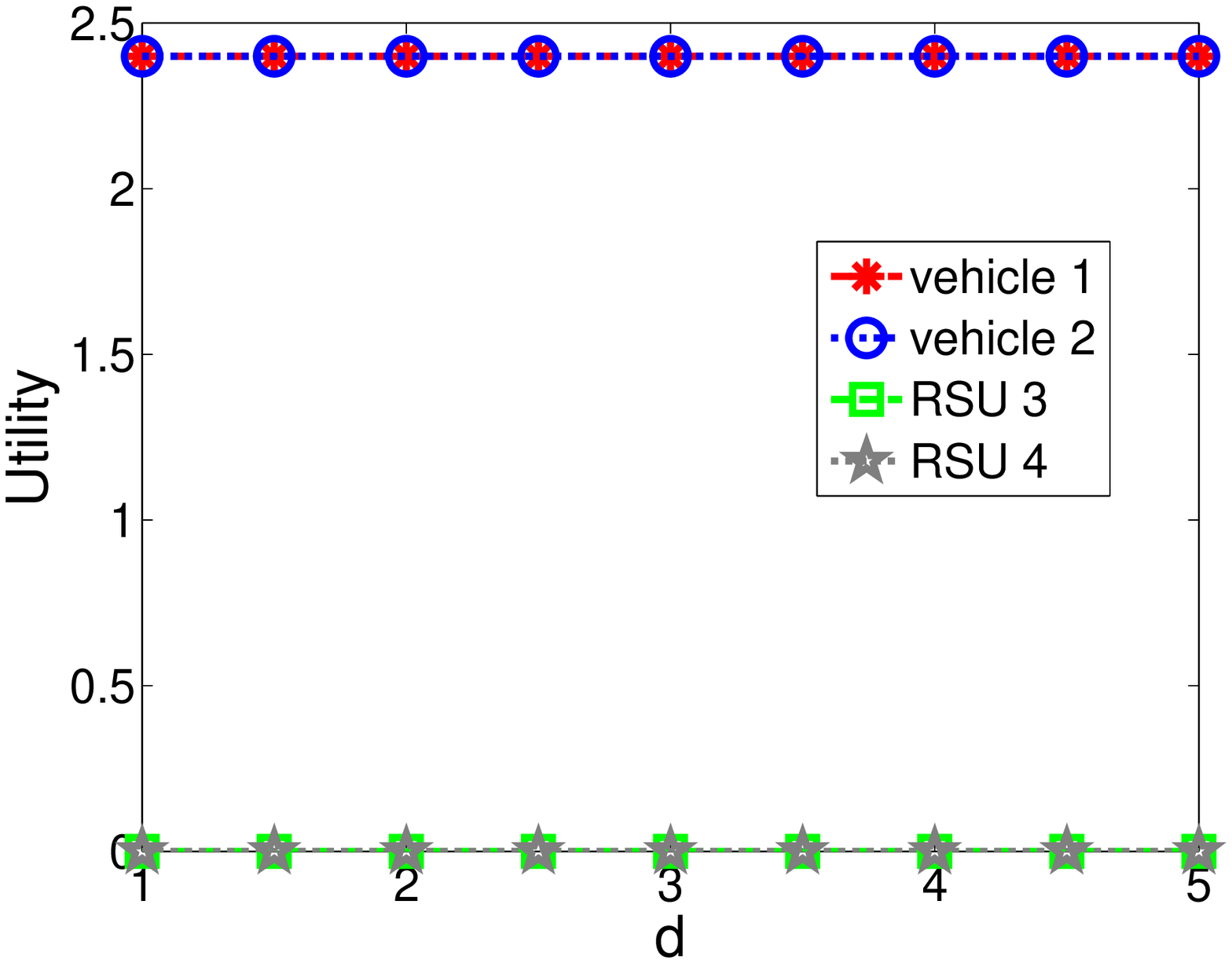}}
\subfigure[ $\mathcal{C}_5$]{\label{C5}\includegraphics[width=1.7in]{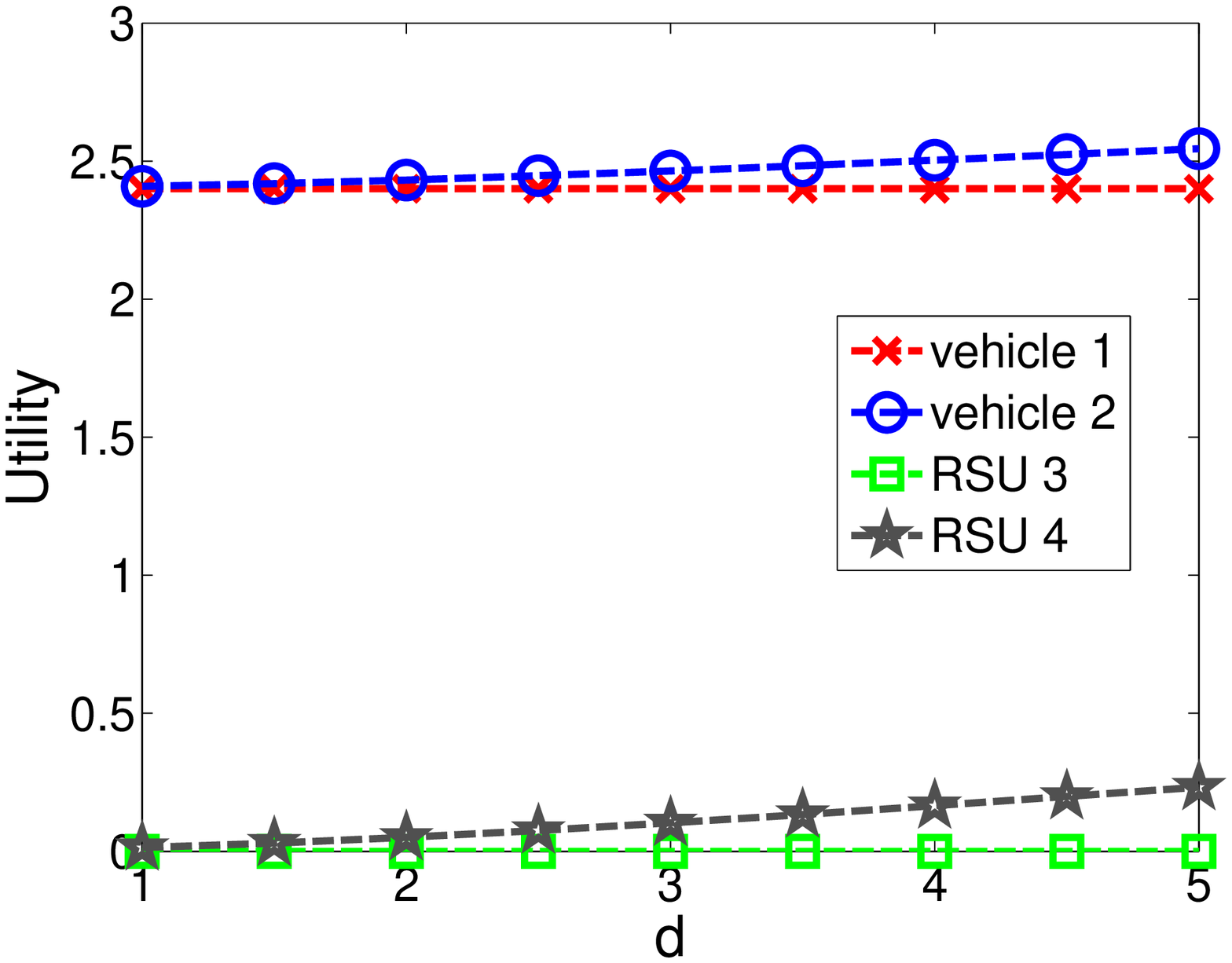}}
\caption{Utility performance in $\mathcal{C}_4$ and $\mathcal{C}_5$}
\end{center}
\end{figure}
\par
\begin{figure}[!t]
\begin{center}
\subfigure[$\mathcal{C}_6$]{\label{C6}\includegraphics[width=1.7in]{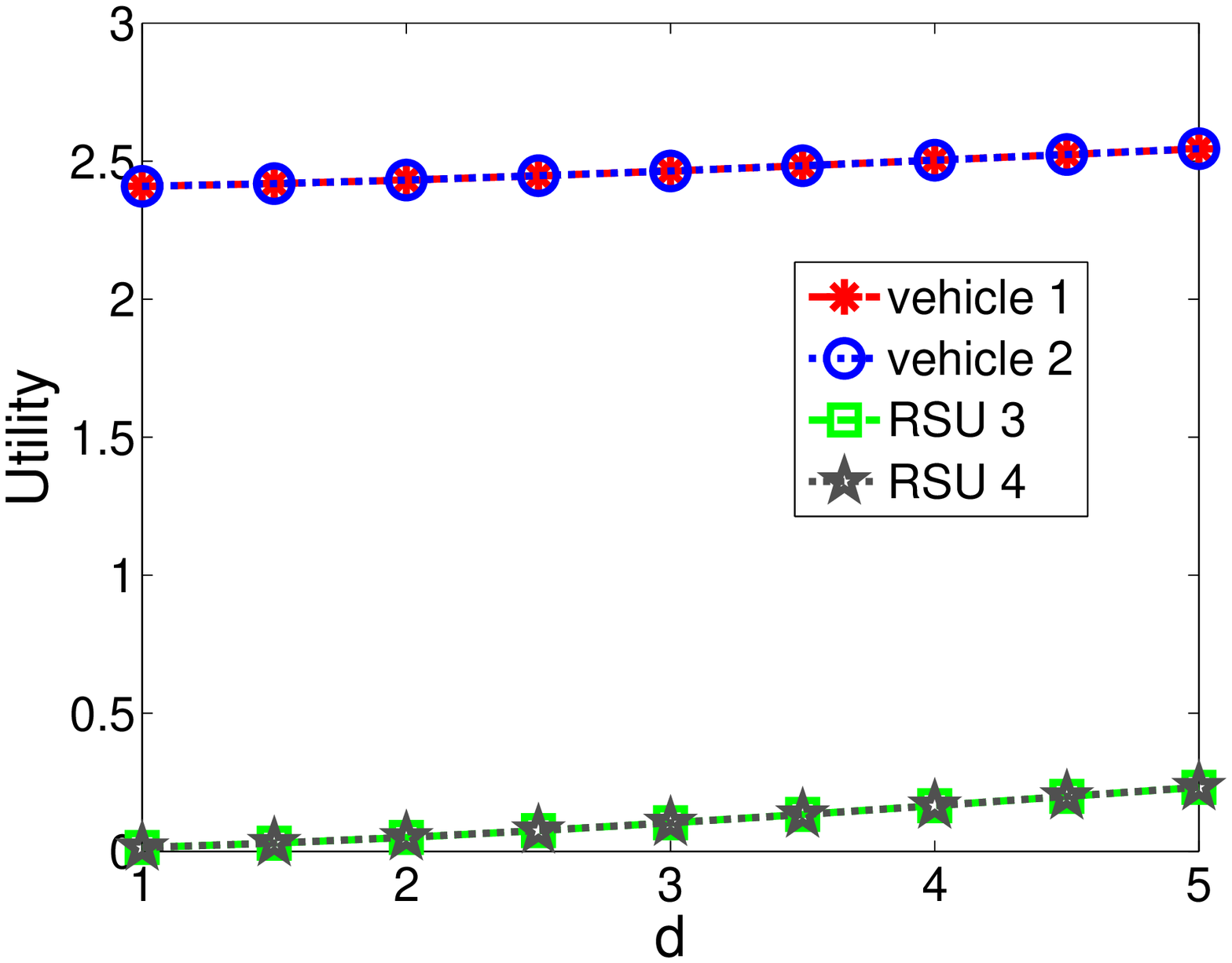}}
\subfigure[$\mathcal{C}_7$]{\label{C7}\includegraphics[width=1.7in]{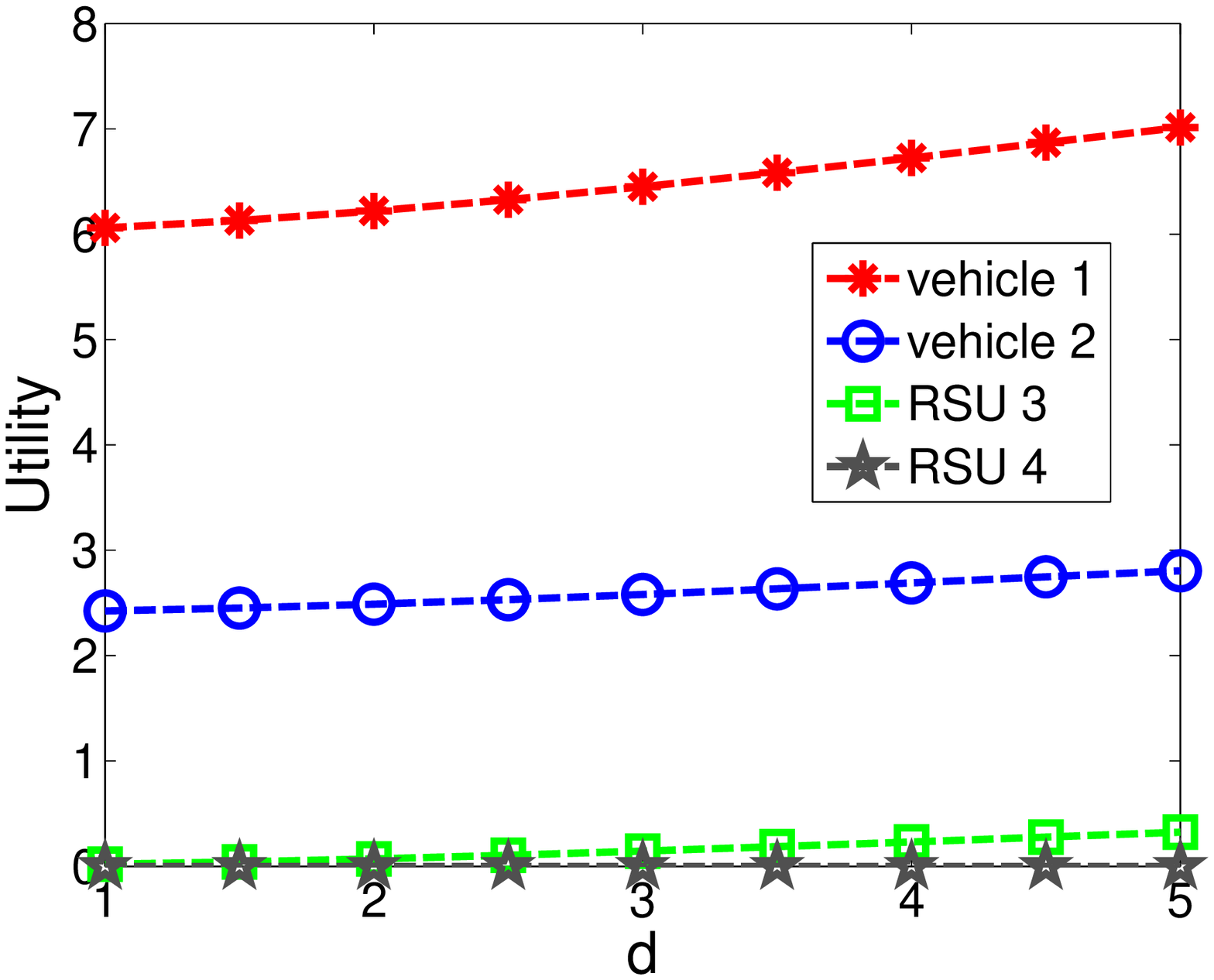}}
\caption{Utility performance in $\mathcal{C}_6$ and $\mathcal{C}_7$}
\end{center}
\end{figure}

Fig. \ref{C2} - Fig. \ref{C6} illustrate the utility performance for $\mathcal{C}_2$ - $\mathcal{C}_7$, respectively.
As compared with Fig. \ref{C1}, the utility performance of vehicle 1 evidently decreases, and the utility performance of vehicle 2, RSU 3, and RSU 4 slightly decreases in Fig. \ref{C2}. Based on the scheduling scheme, the successful transmission probability of RSU 1 is $p_1=0.6$ in $\mathcal{C}_1$. In contrast, it is $p_1 \times (1-p_2)=0.24$ in $\mathcal{C}_2$. That is to say, the successful transmission probability obviously decreases. Thus, the utility decreases evidently. With respect to vehicle 2, the successful transmission probabilities are the same in $\mathcal{C}_1$ and $\mathcal{C}_2$ (i.e., $(1-p_1)\times p_2=0.24$). However, there are no RSUs that are in the same coalition with vehicle 2 in $\mathcal{C}_2$. Then no cooperative transmission can be implemented for vehicle 2, and the utility performance decreases accordingly. In $\mathcal{C}_1$, both vehicle 1 and vehicle 2 may utilize RSU 3 (or RSU 4) for cooperative transmission. In contrast, only vehicle 1 may utilize RSU 3 (or RSU 4) in $\mathcal{C}_2$, i.e., the probability of being utilized as a relay will decrease in $\mathcal{C}_2$.
As being utilized as a relay is profitable in our settings, the utility performance of RSU 3 (RSU 4) decreases.
\par
Comparing Fig. \ref{C3} with Fig. \ref{C4}, it can be observed that the utility performance of vehicle 1
is much better in Fig. \ref{C3}, and the utility performances of the other three nodes (vehicle 2, RSU 3, and RSU 4) are the same. It can be
explained by using (\ref{condition for forming coalition for pure vehicle}). Firstly, $S=\{2,1\}$ with $s_1=2$ and $s_2=1$. Then we have 
\begin{eqnarray}\label{analysis of sim c3 c4}
\left\{
\begin{array}{ll}
\prod\limits_{j \in S\backslash \{s_i\}}\left(1-p_j\right)=1-p_2=0.4 <1, &  i=2;\\
\prod\limits_{j \in S\backslash \{s_i\}}\left(1-p_j\right)\left[\prod\limits_{k =i+1}^{|S|}\left(1-p_{s_k}\right)\right]^{-1}\\
\quad = (1-p_1)[1-p_1]^{-1}=1, &  i=1.
\end{array} \right.
\end{eqnarray}
That is to say, the utility of RSU $s_2=1$ will increase and the utility of RSU $s_1=2$ will
remain the same when they form the coalition $\{2,1\}$.
\par
Finally, we can see that the utility performance in $\mathcal{C}_1$ is better than other coalitional structures and the utility is positive in all coalitional structures. Meanwhile, $\alpha_i=10>0$, $\beta_i=1>0$, and $\gamma_j=\mu_j=1>0$. Thus, the conditions in Observation \ref{sufficient condition for the existence of the core} hold. Applying Observation \ref{sufficient condition for the existence of the core}, we claim that the core of the coalitional game is nonempty under the simulation settings. Furthermore, $\left(u_1(\mathcal{C}_1),u_2(\mathcal{C}_1),u_3(\mathcal{C}_1),u_4(\mathcal{C}_1)\right)$ is in the core.
\section{Conclusion}
Cooperation among vehicles and cooperation between vehicle and RSU in vehicular networks have been studied. We propose the notion of encounter to characterize the relative location between the vehicle and RSU when the vehicle is locomotive.
Utilizing the coalitional game theory and pricing mechanism,
we have formulated a NTU coalitional game to analyze the behaviors of the vehicles and RSUs. Moreover, the stability of the proposed game is studied. A sufficient condition for the non-empty of the core is obtained.
Numerical results for the 2-vehicle and 2-RSU scenario verify the theoretical analysis.


%

%
%
%
\section*{Acknowledgment}

%
This work is partially supported by the National
Basic Research Program of China (973 Program) under
Grants 2013CB336600 and 2012CB316001, the National Nature Science
Foundation (NSF) of China under Grants 60832008, 60902001, and 61021001, US NSF CNS-1117560, CNS-0953377,
ECCS-1028782, CNS-0905556, CNS-1265268, and Qatar National Research
Fund.
\ifCLASSOPTIONcaptionsoff
  \newpage
\fi



%

\end{document}